\documentstyle[preprint,epsfig,aps]{revtex}
\draft
\tighten
\begin{document}

\title{Structure of the proton drip line nucleus $^{17}$F}
\author{K. Krishan$^1$,\ P. Banerjee$^1$\ and R. Bhattacharya$^2$}
\address{$^1$Variable Energy Cyclotron Centre, 1/AF Bidhan Nagar,
Calcutta - 700 064, India\\$ ^2$Gurudas College, Calcutta - 700 054, India}
\date{\today}
\maketitle

\begin{abstract}
The odd  $A$  proton  drip  line  nucleus  $^{17}$F  has  been  studied  in  a
semi-microscopic model which couples proton quasiparticle  motion  to  the
vibrational   motion   of   the  neighbouring  even-even  $^{18}$Ne  core.
The experimentally observed low lying excitation spectrum, electric quadrupole  moment,
magnetic dipole moment,  $B(E1)$ and $B(E2)$ values have been fairly well reproducd.
The calculated rms radius of the first excited state is well reproduced and is
found  to  be  larger  than that of the ground state which agrees with the
experimental observation.
\end{abstract}

\pacs{PACS number(s): 21.60 Ev, 27.20 +j}

\section{Introduction}
With  the advent of radioactive ion beam facilities, new experimental data
concerning exotic nuclei are becoming available. Some of the exotic nuclei
show halo structures which has generated a lot of interest, in  recent  times, among various
theoretical   and   experimental  groups.  Experimental
investigations  have  shown  the   enhancement   of   sub-barrier   fusion
cross-section in reactions involving exotic nuclei \cite{e1,e11}, which is
thought  to be related to the sizes and structure of such nuclei. Thus the
understanding of the structure of such nuclei is of utmost  importance  to the
study of various phenomena involving these nuclei.

It is well known that the very low binding energy of the last unpaired nucleon
is  responsible  for  the large matter radii of one-nucleon halo nuclei. A
number of one-neutron halo nuclei have been identified and  studied  in  recent
times.  In  a similar way, one-proton halo nuclei ought to exist except
for the fact that the  proton  wave  function,  unlike  the  neutron  wave
function, is squeezed due to the Coulomb barrier. The nucleus $^{17}$F may
be  a  good candidate for a proton halo nucleus since the binding energy of the
last  proton  is  only 0.60 MeV. Experimental studies \cite{e2,e3,e4} have
revealed  that  the  spin-parity of the ground  state  of  $^{17}$F  is ${5\over 2}^+$ and has
spectroscopic strength equal to  0.93. The first  excited  state  is  known  to  be
${1\over   2}^+$,   with a dominant   $2s_{1/2}$  single  particle  (s.p.)
configuration having  binding energy equal to 0.1 MeV. However, the root mean square (rms)  radius  of
the  ground  state is 3.78 fm whereas that of the  first excited state, at 500 keV excitation, is
observed to be 5.33 fm \cite{e5}. Considering the  ground  state  and  the
first  excited  state  to  be   pure  single  particle configurations of
1d$_{5/2}$ and 2s$_{1/2}$, respectively, the rms radii of the last proton,
using Harmonic oscillator wave functions, turn out to  be the same,  which 
is
contrary to the experimental findings. This indicates that coupling of the
odd  proton  motion  and  that  of the core may modify the wave function of
$^{17}$F sufficiently to affect the rms radius of the  ground  state  and
that  of  the first  excited  state. The neighbouring even-even cores for this
nucleus are $^{16}$O and $^{18}$Ne. The  first,   $^{16}$O,  is  a
doubly  closed  shell  nucleus  without  any vibrational structure, whereas the second,
$^{18}$Ne, shows a well developed vibrational structure  with  the first  2$^+$
state  at  1.88  MeV  \cite{ne}. As such, the behaviour of the core changes
abruptly if one goes from O to Ne.  Therefore,  the  nature  of the $^{17}$F
nucleus  is  in between the two extremes and the structure of this nucleus
needs  to  be  explored.  Furnstahl  and  Price  \cite{furn},  through   a
relativistic  Hartree  calculation,  found  the rms charge radius to be
2.76 fm and the magnetic dipole moment to be 4.87 $\mu_N$. On  the  other  hand,  the
density  functional  method  of  Lombard  \cite{lomb}  provided  the total
binding energy (B.E. = 126.6  MeV).  However, the  detailed  structure  
information regarding the excitation mechanism is lacking at present.

We  have shown earlier \cite{be11} that the structures of the one-neutron halo
nuclei $^{11}$Be and $^{19}$O could be  explained  satisfactorily  through
the  coupling  of  collective  vibrations  of  the respective even-even 
cores with the motion of the odd valence neutron. In the present work, we have 
shown
that  similar  idea can be applied in explaining the important features of
the excitation spectrum of the proton rich exotic nucleus  $^{17}$F  by  coupling
single  proton  qusiparticle motion to the collective vibrational motion of
the even-even $^{18}$Ne core. In this paper, we present the results of  our
calculations  for the    studies of   the   low   lying   excitation   spectrum  and
the electromagnetic (EM) properties of the $^{17}$F nucleus. The model used 
in
the present calculations is discussed briefly in section II. The method of
calculation is given in section III. Section IV contains the  results  and
discussion and finally the conclusions are given in section V.

\section{MODEL}

The  model  used  in the present calculations is given in our earlier work
\cite{be11,k12}  and  here  we  are  giving  it  just  for  the  sake   of
completeness.  The total Hamiltonian for a coupled system of a quasiparticle
and collective excitation of the core may be written as

\begin{equation}
{ \it H = H_{core} + H_{s.p.} + H_{int}\\}
\end{equation}
where ${\it H_{core}}$ describes the harmonic collective vibrations of the
even-even  core.  {\it  ${H_{s.p.}}$}  describes the motion of the valence
nucleon (-hole) in an effective potential.  The  basis  states,  which  are
eigen  functions  of  ${\it H_0 = H_{core} + H_{s.p.} }$, can be written as
$\{[N_2 V_2 L_2, N_3 V_3 L_3]^L (n l {1/2})^j  \}^{JM}$,  where  $(N_\lambda
V_\lambda  L_\lambda)$  is  the  totally  symmetric  state  of $N_\lambda$
phonons of multipolarity $\lambda$ coupled to angular momentum $L_\lambda$
and $V_\lambda$ represents all the additional  quantum  numbers  necessary
for  complete  specification of the state; $L_2$ and $L_3$ are coupled to
the angular momentum $L$ of the core which is then coupled to $j$  of  the
quasiparticle  to  form  the  resultant  $J$.  ($  n  l 1/2$) represents a
quasiparticle state  having  radial  quantum  number  ${\it  n}$,  orbital
qauntum  number  ${ l}$ and spin ${ 1/2}$. The eigen values of ${\it H_0}$
are the sum of the single particle  energies  $\epsilon_p$  and  the  core
energies. The core particle interaction is given by

\begin{equation}   H_{int}   =
\sum_{\lambda}^{2,3}   K_{\lambda}(r)   \sum_{\mu   =   -\lambda}^\lambda
\alpha_{\lambda\mu}^* Y_{\lambda\mu} (\theta_p, \phi_p), \end{equation}
$Y_{\lambda\mu}$ are the spherical harmonics corresponding to the particle
coordinates.  The  collective  coordinates  $\alpha_{\lambda\mu}$  can  be
expressed as a linear  combination  of the phonon  creation  and  destruction
operators  $b^+,  \  b$ respectively, for the core. The matrix elements of
the interaction Hamiltonian, after taking care of the pairing effects, are
given by

\begin{eqnarray}
&&<\{[N_2 V_2 L_2, N_3 V_3 L_3]^L (nl1/2)^j \}^{JM}|H_{int}| \nonumber\\
&&\hspace{2cm} \{[N'_2 V'_2 L'_2, N'_3 V'_3 L'_3]^{L'}(n'l'1/2)^{j'} \}^{JM}>\nonumber\\
&&= \sum_\lambda^{2,3} X_\lambda <nl||[{({{m \omega_0}\over {\hbar}})}^{1/2}
r_p]^{\lambda}||n'l'> (-1)^{J-1/2+{\lambda}}\nonumber\\
 &&\times {1\over 2} \times
\left\{ [L] [L'] [j] [j'] \right\}^{1\over 2} \times
\left ({1 + (-1)^{l+l'+\lambda}\over 2}\right)\nonumber\\
&&\times\left(\delta_{\lambda,2} + \delta_{\lambda,3} (-1)^{L+L'+L_3+L'_3}
\right)
\left(
\begin{array}{ccc}
j'& \lambda& j\\
1/2& 0& -1/2
\end{array} \right)\nonumber\\
&&\times\left\{
\begin{array}{ccc}
L& L'& \lambda\\
L'_{\lambda}& L_{\lambda}& L_{\eta}
\end{array}\right\}
\times (U'_j U_j - V'_j V_j) \delta_{N_\eta N_{\eta'}} \delta_{L_{\eta}
L_{\eta'}} \delta_{V_{\eta} V_{\eta'}} \nonumber\\
&&\times \Bigl[ \delta_{N_{\lambda} N'_{\lambda -1}} \times (-1)^{L_2 + L_3}
\times <N_{\lambda} L_{\lambda}|| b_{\lambda}^+ || N_{\lambda'}
L_{\lambda'}>\nonumber\\
&& + \delta_{N'_{\lambda} N_{\lambda+1}} \times (-1)^{L'_2 + L'_3}
<N'_{\lambda} L'_{\lambda}|| b_{\lambda}^+ ||
N_{\lambda} L_{\lambda}>\Bigr]. \nonumber
\end{eqnarray}
In the above, $X_2$  and  $X_3$  are  the  quadrupole and octupole interaction strengths
respectively,  and  $\eta  =  2$  when  $\lambda  =  3$  and  vice  versa.
$X_\lambda$'s are given by

\begin{equation}
X_{\lambda} = K_\lambda \sqrt {2(\lambda +  1)}
\sqrt{  {\hbar  \omega_\lambda}  \over  {2  \pi  C_\lambda}},
\label{ki}
\end{equation}
where $K_\lambda$ are the average values of the radial integrals. The quantities
$V_j$'s and $U_j$'s are the occupation and non-occupation probabilities of
the  respective  single  particle states. Diagonalising the Hamiltonian of
the coupled system in the basis space of ${\it H_0}$, one obtains the eigen
values and the eigen vectors.  The  eigen vectors, thus obtained, are used to
calculate the static electric quadrupole and magnetic dipole moments, the electric and  the magnetic
multipole  transitions. The expressions for the static magnetic dipole and
the electric multipole operators for the coupled system are \cite{bo53}

\begin{equation}
 O(M1,\mu) = {(3/4\pi )}^{1/2}\left(g_{l}l_{\mu }+g_{s}s_{\mu }+g_{R}R \right), \\
\label{mag}
\end{equation}
\begin{equation}
O(E\lambda ,\mu ) = {3 \over {4\pi}}  Ze {R_0}^\lambda \
{({{\hbar \omega _\lambda} \over {2C_\lambda}})^{1\over2}} \
(b^{+ }_{\lambda ,\mu } + (-)^{\mu }b_{\lambda ,-\mu }) + (e_p + {Ze \over A^2})  \ r_{p}^{\lambda }
 \ Y_{\lambda ,\mu }(\theta ,\phi )
\label{QM}
\end{equation}

\noindent where  $Z$,  $A$  and  $R_{0}$  are the charge, mass and the radius of the
core, respectively. The quantity $C_{\lambda}$  refers  to  the  stiffness
constant  of  the vibrating core. The magnetic dipole moment $\mu$ and the
electric quadrupole moment (QM) $Q$ of a state $  |  J,  M,  \alpha>$  are
defined as

\begin{equation}
 \mu  = < J, M=J, \alpha| \ {({4\pi \over 3})}^{1/2} O(M1, 0) \ | J, M=J,\alpha>\\
\end{equation}
and
\begin{equation}
Q =  < J, M=J, \alpha| \ {({16\pi \over 5})}^{1/2} O(E2, 0) \ | J, M=J,\alpha>\\.
\end{equation}
The reduced  eletromagnetic  transitions are given as
\begin{equation}
 B(E\lambda : J_{i} \to J_{f})  = <J_{f} || \ O (E\lambda ) \  || J_{i}>^{2} {(2J_{i} + 1)}^{-1}
\end{equation}

The  detailed  expressions, after carrying out angular momentum algebra, are
given in Heyde and  Brussaard  \cite{hyde}  except  for  a  multiplicative
factor in the particle part invoving $V_j$'s and $U_j$'s. The
rms radius of a state is obtained by taking the expectation value of the
operator  $r_p^2$,  in  the respective state. The core contribution to the rms
radius  is  either  taken from the experiment, if available. Otherwise, it  is taken as
${R_0}^{2} = {(1.2 A^{1/3})}^2$.

\section{Method of Calculation}

There  are  several  parameters in this calculation viz. (i) the quasiparticle
energies $\epsilon_j$, (ii)  the  $U_j  (V_j)  $  factors  and  (iii)  the
interaction  coupling  strengths $X_2$ and $X_3$. The vibrational energies
for the $^{18}$Ne core are taken from the experimental data \cite{ne}. The
experimental excitation spectrum of $^{18}$Ne  shows  the  behavior  of  a
vibrator.  The  first  $2^{+}$ excited state occurring at 1.887 MeV can be
identified as one quadrupole phonon excited state. The second $0^{+}$ ,
$2^{+}$ states and the first $4^+$ state occurring at 3.576 MeV, 3.616 MeV and
3.376 MeV, respectively, can be taken as the members of the multiplet of  two  quadrupole
phonon  excitations.  The  calculations  have  been restricted to only two
quadrupole phonon space as the calculated spectra have been  found  to  be
insensitive to the inclusion of octupole phonons. Therefore, $X_3$ is kept
equal  to zero. The single particle space, in the present calculations, is
confined to $1d_{5/2}$, $2s_{1/2}$ and $1d_{3/2}$ orbitals for the positive  parity
states  and  $1p_{3/2}$, $1p_{1/2}$  orbitals for  the negative parity states.
The $U_j (V_j) $  factors  for  the $1d_{5/2}$,  $2s_{1/2}$  and
$1d_{3/2}$ single particle states are available from the ($\alpha ,t$) and
($^3He,d$)   reaction   studies  \cite{yasu,e3}  .  However,  the experimental
spectroscopic information for the $1p_{3/2}$  and  $1p_{1/2}$  single  particle
states is not very clear. From the ($d,^3{He}$) reaction, Firestone et
al. \cite{fire} overestimated the  strengths for the $1p_{3/2}$ and $1p_{1/2}$ 
single particle states by $30 - 40 \%$ and, to the best of our knowledge, no other
data are available. Therefore, we have taken them as free parameters.  The
$U_j  (V_j)  $  factors  used  in the present calculation to reproduce the
experimental data  are  given  in  Table  I.  The  relative  quasiparticle
energies,  $\epsilon_j$,  with respect to the lowest single particle state
in the  conventional  shell  model  basis  and  the  interaction  coupling
strength $X_2$ are treated as free parameters. These are varied to fit the
experimental  level  scheme and  the  spectroscopic  factors. 
 The set of best fit parameters used in the present calculations  are given
in Table \ref{parameter}.

The  wave function of a state with angular momentum $J$ with $z$-component
$M$ and energy $E^{\alpha}$ ($\alpha$ distinguishes between states of same
spin and parity) is given by

\begin{equation}
|E^{\alpha} , JM> = \sum_{N_2,L_2,j} C_{\alpha} (N_2 L_2,j,J) |N_2 L_2,j;J M>,
\end{equation}

and the spectroscopic factor is defined as

\begin{equation}
S^\alpha (l,j=J) = U^2_j | C_\alpha (00,j,J) |^2 .
\end{equation}

\section{Results and Discussion}

\subsection{Even parity states }

For the even parity states, the odd quasiparticle motion was restricted to be in the
$1d_{5\over  2}$,  $2s_{1\over  2}$  and  $1d_{3\over  2}$  orbitals.  The
collective states of $^{18}$Ne, used in the present calculation, to  which
the   quasiparticle  couples,  were  taken  from  the
experimental  data  \cite{ne}.  The  calculated  low  lying   levels   and
the spectroscopic factors are shown in Fig.1, along with the experimental data,
\cite{till}, for the sake of comparison. It is seen from  this figure  that
the present calculation is able to reproduce  the positive parity
level  spectrum quite well.  The present calculation predicts two weak ${5\over 2}^+$ and
${1\over  2}^+$  states, not  observed experimentally, at  $\sim$2  MeV  excitation  with 
 the spectroscopic strengths  0.05 and 0.06, respectively (not shown in Fig. 1).
. The calculated spectrum is found to
have  three  ${3\over  2}^+$  states  lying  between  4.50  and  4.75  MeV
excitation   with  single  particle  strengths equal to 0.05,   0.51   and   0.07,
respectively. The state at 4.63 MeV with a spectroscopic strength of 0.51  may
correspond  to  the  experimentally observed ${3\over 2}^+$ state at 5 MeV excitation
with spectroscopic strength 0.54. Whereas, the calculated state at 4.75  MeV
may  correspond to the observed state at 5.82 MeV. The calculated ${1\over
2}^+$ and ${3\over 2}^+$ states at 6.83 MeV and 6.84 MeV may correspond to
the observed levels at 6.56 MeV and 7.36 MeV, respectively. It is conjectured
that the experimentally observed ${5\over 2}^+$ state at 8.07  MeV  may  be
associated  with  the  calculated  state  at 8.20 MeV.

There  is  an  overall  good  agreement  between  the  calculated and the
experimental level energies and the spectroscopic factors. However, almost all
the calculated ${3\over 2}^+$ states are depressed in energy by 500 keV as
compared to the correponding experimentally observed states. The observed $1d_{5\over
2}$ single particle strength is  found to be almost totally concentrated on the ${5\over 2}^+_1$  
state  though  the  present calculation predicts a ${5\over 2}^+$ state at 2 MeV
with a very small spectroscopic strength. A similar  situation  is  observed
with the $1d_{3\over 2}$ single particle strength distributions. The available
$1d_{3\over 2}$ single particle strength is found to be fragmented amongst
three states lying between 4.50 MeV and 4.75 MeV, the major part of strength
being concentrated on the state at 4.63 MeV excitation.

\subsection {Odd parity states }

For generation of the odd parity states, the single particle space considered
in  the  present calculation is spanned by $1p_{1\over 2}$ and $1p_{3\over
2}$ states. The quadrupole interaction coupling strength  $X_2$  was  kept
identical to that  used in the case of the even parity states and the single particle
energies $\epsilon_{1p_{1\over 2}}$ and  $\epsilon_{1p_{3\over  2}}$  were
treated as free parameters. The calculated level spectrum of the low lying odd
parity  states  is  shown  in  Fig.  2  alongwith  the  experimental one
\cite{till}. It is  seen  from  this figure  that  there  is  an  overall  good
agreement  between  the  calculated  and the experimental  spectra except for
the ${5\over 2}^-$ states which are overpredicted by $\sim$ 1  MeV.  The  calculation
predicts  that  almost  the  total  available  $1p_{1\over  2}$ single particle
strength is concentrated on the ${1\over 2}^-$ state at 3.186 MeV. The $1p_{3\over  2}$  single
particle  strength is predicted to be fragmented into three ${3\over 2}^-$
states lying at 4.74 MeV, 5.20 MeV and 7.33 MeV with  major  part  of  the
strength  going  to the state at 4.74 MeV. The structure of the calculated
${5\over 2}^-$ state at 4.86 MeV is found to be mainly consisting of a $2^+$
collective state coupled to the $1p_{1\over  2}$  single  particle  orbital.
Similar  are  the  structures  of the ${7\over 2}^-$ and ${9\over 2}^-$ states.
However,  no  experimental  data  are  available  for  the   spectroscopic
strengths for the odd parity states of $^{17}$F to compare with.

\subsection {RMS radii, EM moments and Transition probabilities }

The  stringent  test  of the wave functions  obtained  by  diagonalising the
Hamiltonian is to see whether they are able to  reproduce the electromagnetic
moments and the multipole  electromagnetic transition probabilities. However, in
the case of the exotic nuclei an important aspect of structure  study  is  the
reproduction  of  the  experimentally  observed rms radius of proton or
neutron distribution which provides a measure of the proton or  neutron
halo. The rms radii of the ground
state and the excited states of $^{17}$F contains two parts - (i) from the core
and  (ii) from the extra quasiparticle.
In order to evaluate the proton rms radii of these two
 states, initially, we performed our calculation with the Harmonic
Oscillator (H.O.) wave functions to extract the contribution from the
quasiparticle part.  The calculated values of the proton rms  radius  for  the  ${5
\over  2}^+$  ground state and the ${1 \over 2}^+$ first excited state were found
to be 4.01 fm and 4.15 fm, respectively. Though the experimental trend for
the rms radii for these two  states  is  reproduced,
the  calculated  values  are  not in agreement with  the
 experimentally observed values  of  3.78  fm  and  5.33  fm,
respectively  \cite{e5}. This discrepancy may be attributed to the 
fact that the H.O. wave functions do not carry any effect of the 
low binding enegy of the last valence nucleon. This behaviour of 
the H.O. wave functions, through the radial integrals, consequently
affects the calculated values of the electric quadrupole moment and
the electromagnetic transition rates (Eq.\ref {QM}).

To overcome this difficulty, we have used a generalised average 
one-body potential of Woods-Saxon type \cite{cal}  to generate the s.p. wave functions.
Keeping all the parameters fixed,  the depth  $V_0$  and  the  radius
parameter $r_0$ were varied to reproduce the experimentally observed binding energies,
0.600  MeV  and  0.105  MeV  of the $1d_{5\over  2}$ and $2s_{1\over 2}$ s.p.
states, respectively. The parameters of the single particle potential  are
given  in  Table  \ref{wspot}.  These W-S wave functions could be used for
calculating the rms radius  and  the  electromagnetic  properties  of  the
states  of  $^{17}$F  but  for  the wave function of the $1d_{3\over 2}$ 
state, which is unbound.

 In order to circumvent this  problem  to  get  the  wave
function of the unbound $1d_{3\over 2}$ s.p. state, we developed the W-S wave
function in the H.O. basis (WSHO). This was done by digonalising  the  W-S
potential (Table \ref{wspot}) for each ($ l,j$) state in H.O. basis states
($n,l,j$)  with  nodal  quantum  number $ n$=10 and properly picking up the
$1d_{5\over 2}$, $2s_{1\over 2}$, $1d_{3\over  2}$,  $1p_{3\over  2}$  and
$1p_{1\over  2}$  states.  The  $<r^2>$  values  for  the even parity s.p.
states, thus developed, are given in Table \ref{integral}. From the  table
it  is  observed  that $<r^2>$ values calculated with the W-S wave function in
the H.O. basis are smaller than those calculated with the W-S wave functions
and quite larger than those calculated  with the H.O.  wave  functions. The
calculated values of the rms radii of $^{17}$F using the quasiparticle wave
functions WSHO are given in Table \ref{qmr}. For the first excited  state,
the  calculated  value  of  the rms  radius  agrees  quite  well with the
experimental one. However, for the ground state the  calculated  value  of
the rms radius overestimates the corresponding experimental value.

The same quasiparticle wave functions, as used in the calculation of the rms
radii, have been used to calculate the electric quadrupole  moment  of  the  ground
state and the EM transition probabilities among the various states of $^{17}$F
obtained  earlier.  Effective proton charge, ${(e_p)}_{eff} = 1.5 \ e$ and
effective charge of the core, $Z_{eff} = 0.2 \ e $ have been used in  the
present  calculation  ($e$ being the charge of proton). However, we made a
rough estimate of the value of $Z_{eff}$ (defined as $Z {(\hbar \omega_2/2
C_2)}^{1/2}$) from the value of the quadrupole interaction strength parameter
$X_2$ ( see Eq. \ref{ki} ) used earlier. Taking value of $\hbar \omega_2$,
the  quadrupole  phonon  energy $\approx$ 2 MeV, and of $<K>$, the average
radial integral $\approx$ 40 MeV, the  calculated  value  of  $Z_{eff}$  is
found to be equal to 0.20 $e$, almost same as the one used in the calculations
of the QM and the EM transition probabilities. The calculated values of the 
QM of
the ground state and the reduced transition probabilities $B(E2)$ and $B(E1)$  are
given  in  Tables  \ref{qmr}  and  \ref{BE} along with their corresponding
experimentally measured values \cite{till}. For evaluation of the magnetic
dipole moment, the values of $g_s$, $g_l$ and $g_R$ were  5.585,  1.0  and
1.0,  respectively  \cite{hydb}.  The  calculated values of the electric quadrupole
moment and the magnetic dipole moment are 0.108 eb and 4.708 $\mu_N$, respectively
and  they compare very well with the experimental values of 0.10 $\pm$0.02
eb  and 4.72130 $\pm$ 0.00025 $\mu_N$, respectively (see Table \ref{qmr}).
The  calculation is able to reproduce well the observed reduced transition
probabilities, $B(E2)$ and $B(E1)$ (Table \ref{BE}). However,  the  calculated
$B(E1)$  value  between the ${1\over2}_1^-$ and ${1\over2}_1^+$ states underestimate
the experimental value by an order of magnitude.

\section{Summary and Conclusion}

The low lying excitation spectrum of the proton drip line nucleus $^{17}$F
has been studied in the quasiparticle-core coupling model  in  which  the
motion of the odd quasiparticle is coupled with the neighbouring even-even
vibrating core. The core excitation energies, which take into account all
the correlations arising out of the many particle core system, were  taken
from  the experiment. The relative single particle energies and the quadrupole
interaction  strength  were treated as parameters. The calculated energies
and  the  spectroscopic  factors for both the even and the odd parity states
agree   quite  well  with  the  corresponding  experimental  results.  The
calculated  values  of  the  rms  radii  of the ground state and the first
excited  state  using  single  particle  H.O.  wave  functions  follow the
experimental  trend  but  are grossly underestimated (particularly the rms
radius of the first excited state). However, the value of the rms radius of
the first excited state is well reproduced by using the WSHO single particle wave
functions. The calculated values of the electric quadrupole moment of the ground  state
and the reduced  transition  probabilities using the WSHO wave functions 
compare
quite well with the corresponding experimental values. The calculated value of
the ground state  magnetic  dipole moment  agrees  very  well  with  that  of  the
experimental one.

In  conclusion,  it  may  be  inferred that within a simple 
quasiparticle-vibration coupling  model, one can very well reproduce the 
experimental observables associated with the structure of the light proton 
drip line nucleus $^{17}$F without going  into  sophisticated calculations. 

\section{acknowledgements}

The  authors  acknowledge  fruitful  discussions  with   S. Sen. One of the
authors  (P.B.)  thanks  Bikash  Sinha     for   moral support
 and J. N. De for his encouragement during the work.

\begin{figure}
\begin{center}\mbox{\epsfig{file=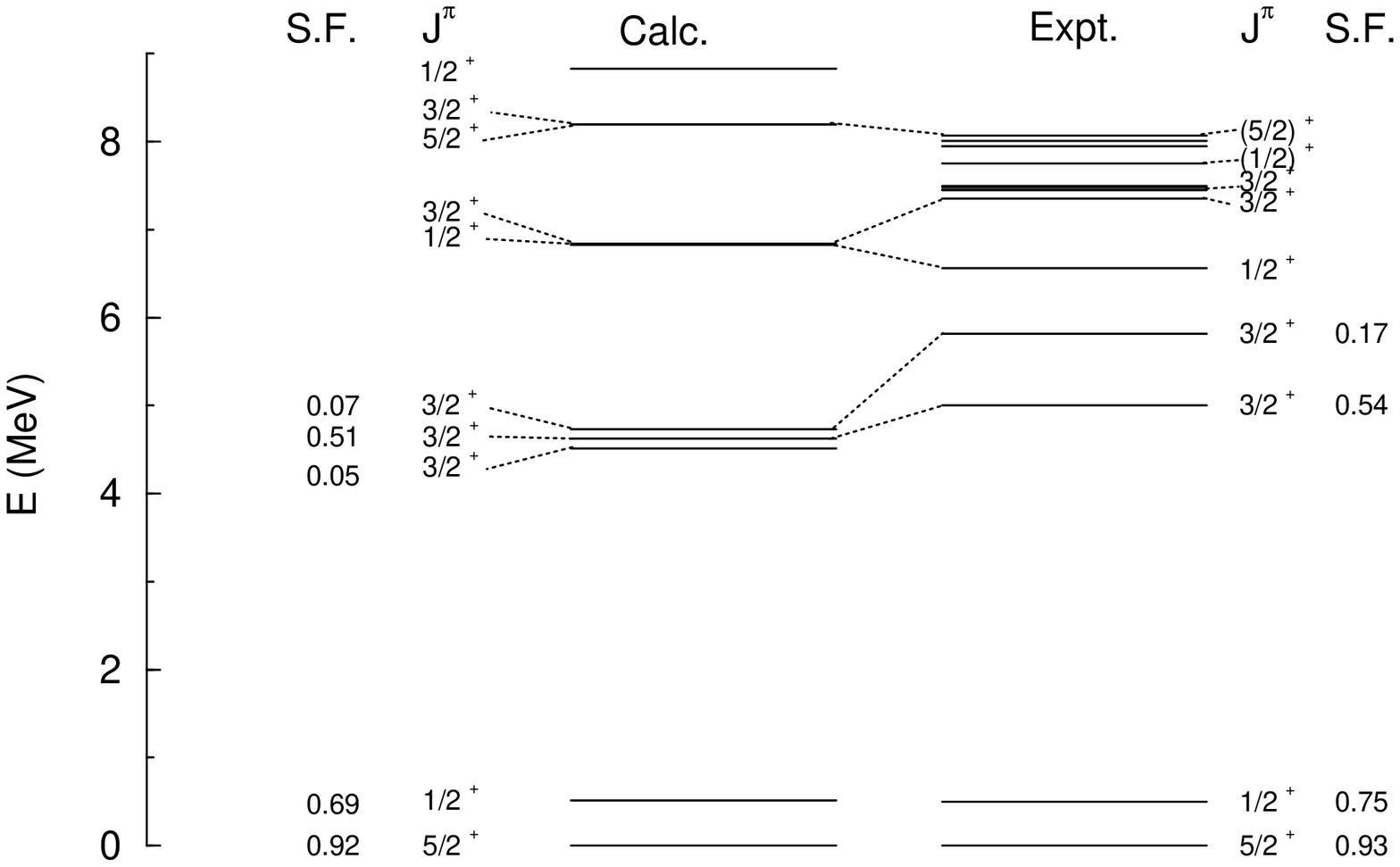,height=10cm}}\end{center}
\caption{  Calculated  and  experimental   even parity levels of  $^{17}$F.  The
excitation energies, spin-parities and spectroscopic factors  of  the  levels
are shown.}
\label{fig:figa}
\end{figure}

\newpage

\begin{figure}
\begin{center}\mbox{\epsfig{file=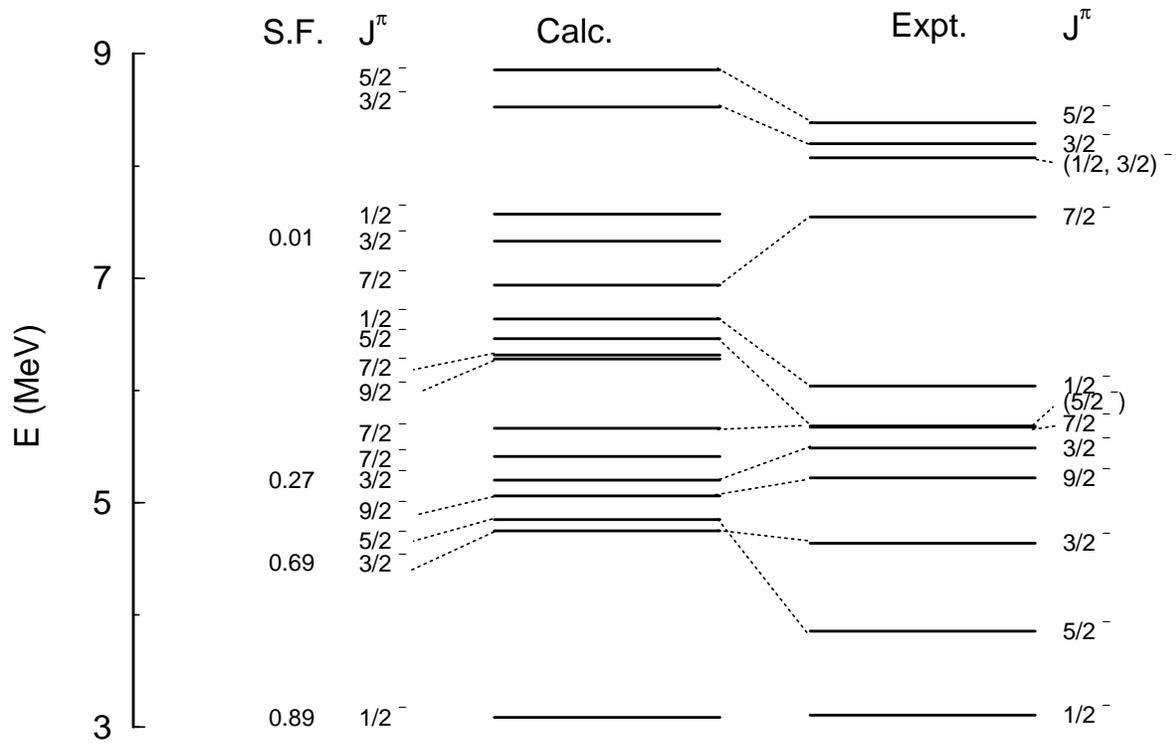,height=10cm}}\end{center}
\caption{ Same as in fig. 1 for odd parity levels  of $^{17}$F.}
\label{fig:figb}
\end{figure}

\begin{table}
\caption{Parameters of the Calculation}
\begin{tabular}{cccccccccccc}
\multicolumn{5}{c}{$U_j$}&\multicolumn{5}{c}{$\epsilon_j$ \ (MeV)}&
$X_2$\\
$1d_{5\over2}$&$2s_{1\over2}$&$1d_{3\over2}$&$1p_{3\over2}$&$1p_{1\over2}$
&$1d_{5\over2}$&$2s_{1\over2}$&$1d_{3\over2}$&$1p_{3\over2}$&$1p_{1\over2}$\\
\tableline
0.985&0.866&0.800&0.142&0.313&0.0&0.550&4.500&4.800&3.00&0.5\\

\end{tabular}
\label{parameter}
\end{table}

\begin{table}
\caption{ Parameters of Woods-Saxon s.p. potential}
\begin{tabular}{ccccccc}
$V_0$&$r_0$&$a_0$&$V_s$&$r_s$&$a_s$\\
\tableline
49.091&1.404&0.685&23.937&0.819&1.060\\

\end{tabular}
\label{wspot}
\end{table}

\begin{table}
\caption{Mean square radius for single particle orbitals}
\begin{tabular}{ccccc}
$<r^2>$&W-S wave fuction in H.O.Basis& W.S wave function\tablenote{Parameters
of Table 1}&H.O. wave function\\
\tableline
$1d_{5\over2}$&14.154&16.012&8.819\\
$2s_{1\over2}$&19.342&29.925&8.819\\
$1d_{3\over2}$&17.690&&8.819\\

\end{tabular}
\label{integral}
\end{table}

\begin{table}
\caption{Electromagnetic moments and rms radius}
\begin{tabular}{cccccccc}
$J^{\pi}$&\multicolumn{2}{c}{Quadrupole Moment (eb)}&\multicolumn{2}{c}
{Magnetic Moment($\mu_N$)}&\multicolumn{2}{c}{rms radius (fm)}\\
&Calc.&Expt.&Calc.&Expt.&Calc.&Expt.\\
\tableline
${5\over2}^+$&0.108&0.10$\pm$0.02&4.7080&4.72130$\pm$0.00025&4.83&3.78\\
${1\over2}^+$&&&2.7706&&5.34&5.33\\

\end{tabular}
\label{qmr}
\end{table}

\begin{table}
\caption{Reduced Electric transitions probabilities $B(E2)$ and $B(E1)$}
\begin{tabular}{cccccccc}
$J_i^\pi$&$J_f^\pi$&$E_i$(MeV)&$E_f$(MeV)& Mult.& Calc. (W.u.)& Expt. (W.u.)\\
\tableline
${1\over2}^+$&${5\over2}^+$&0.517&0.0&E2&21.9&25.0$\pm0.5$\\
${1\over2}^-$&${1\over2}^+$&3.087&0.517&E1&0.2$\times10^{-3}$&(1.5$\pm$0.3)$\times10^{-3}$\\
${5\over2}^-$&${5\over2}^+$&4.848&0.0&E1&6.1$\times10^{-3}$&(4.3$\pm$0.8)$\times10^{-3}$\\

\end{tabular}
\label{BE}
\end{table}
\end{document}